# Rate-Utility Optimized Streaming of Volumetric Media for Augmented Reality

*Jounsup Park, Philip A. Chou, and Jenq-Neng Hwang*[1]

*Abstract* – Volumetric media, popularly known as holograms, need to be delivered to users using both on-demand and live streaming, for new augmented reality (AR) and virtual reality (VR) experiences. As in video streaming, hologram streaming must support network adaptivity and fast startup, but must also moderate large bandwidths, multiple simultaneously streaming objects, and frequent user interaction, which requires low delay. In this paper, we introduce the first system to our knowledge designed specifically for streaming volumetric media. The system reduces bandwidth by introducing 3D tiles, and culling them or reducing their level of detail depending on their relation to the user's view frustum and distance to the user. Our system reduces latency by introducing a window-based buffer, which in contrast to a queue-based buffer allows insertions near the head of the buffer rather than only at the tail of the buffer, to respond quickly to user interaction. To allocate bits between different tiles across multiple objects, we introduce a simple greedy yet provably optimal algorithm for rate-utility optimization. We introduce utility measures based not only on the underlying quality of the representation, but on the level of detail relative to the user's viewpoint and device resolution. Simulation results show that the proposed algorithm provides superior quality compared to existing video-streaming approaches adapted to hologram streaming, in terms of utility and user experience over variable, throughput-constrained networks.

## I. INTRODUCTION

From the moment the first commercial systems for streaming video on demand over the Internet were released in 1997 until the present, streaming video on demand has grown exponentially along several dimensions. The first systems streamed 176x144 (QCIF) video at 15 frames per second (fps), over 56 Kbps modems, compressed to 40 Kbps. Today, 1920x1080 (Full HD) video at 30 fps is regularly streamed to broadband users at 20 Mbps, thanks to a doubling of the last mile bandwidth to the consumer every 2.5 years, and a doubling of the compression quality every 10 years, or faster. During the same period, the Internet backbone capacity has increased 40% annually to about 300 Tbps, not including the proprietary networks of content providers such as Amazon, Facebook, Google, and Microsoft [1]. Cache-based content distribution networks, originally developed for web traffic, were leveraged by basing streaming video protocols on HTTP, and making video servers look like stateless web servers, both for on-demand and live streaming. The combination of these developments has positioned over-the-top (OTT) streaming video on demand to become the dominant means for consuming video, in any form, worldwide. The upcoming deployment of 5G will cement this position by commanding mobile platforms as well.

While OTT live and on-demand streaming video are on their way to becoming the dominant way of delivering video, new forms of immersive media have recently been born, offering experiences well beyond ordinary video. Such new forms of immersive media include *spherical video* for virtual reality (VR), and *volumetric media*, popularly known as *holograms*, for augmented reality (AR) as well as VR. These new media will also be streamed live and on demand.

Spherical (or 360° or omnidirectional) video has a number of variants, including simple monoscopic spherical video with two degrees of freedom (DOF), allowing the user to change the azimuth and elevation of the field of view, suitable for web browsers; 3DOF video, allowing changing yaw, pitch, and roll with correct stereo parallax, more suitable for head mounted displays; 3DOF+ video additionally allowing short translational motion due to the head pivot to reduce motion sickness; and omnidirectional 6DOF video allowing a few steps of translational motion. Of these, the only one with any significant deployment is the simple monoscopic 2DOF spherical video. Essentially, this is ordinary video projected onto a sphere, with the user at the center. The sphere, at $4\pi$ steradians, is about 200 times the steradians of view required to see a picture placed at a typical distance of four times the picture diagonal. Thus, the bandwidth of simple monoscopic 2DOF spherical video could be over two orders of magnitude higher than that of ordinary video. Fortunately, though the user may turn her gaze in any direction, her view is foveated. This allows the spherical video to be streamed more efficiently by attempting to stream high resolution only where the user is looking, and low resolution elsewhere. This is an approach taken in several works [15]-[20] For example, the spherical video can be partitioned into tiles, and the tiles can be streamed with different resolutions depending on where the user is looking.

---

[1] J. Park is with the University of Washington (jsup517@uw.edu). P. A. Chou is with 8i Labs Inc. (phil@8i.com) J.-N. Hwang is with the University of Washington (hwang@uw.edu). This work was done while the first author was at 8i Labs, Inc.

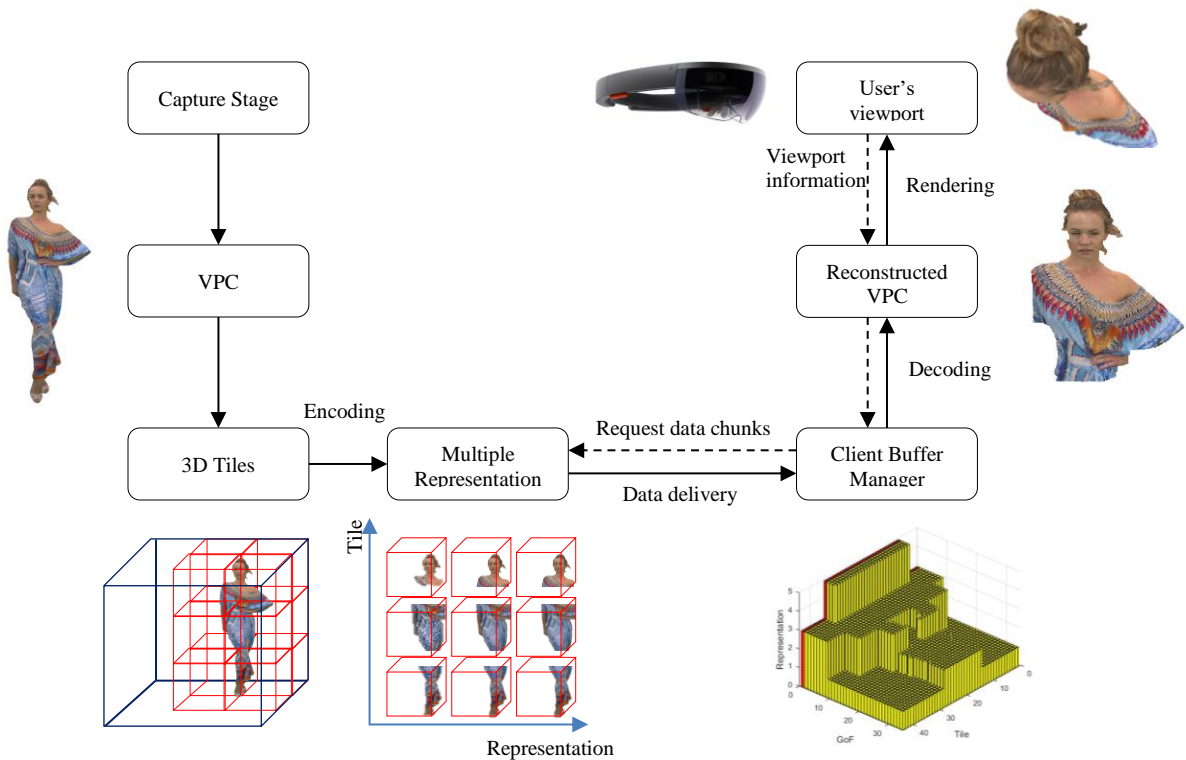

Fig 1. AR Streaming System

Just like streaming ordinary video, streaming spherical video must be *network-adaptive*, that is, adaptive to network variability. For this aspect of streaming spherical video, the same network-adaptive techniques used for streaming ordinary video can be used: buffering, monitoring and predicting the state of the network, and selecting representations at the appropriate bit rates to stream in order to maximize the quality perceived by the user.

The main new challenge that spherical video presents to streaming media is the higher level of *user interactivity* required for spherical video than for ordinary video. For ordinary video, user interactivity is usually limited to starting, stopping, and perhaps fast forwarding the video. For these user interactions, only starting has had a significant impact on the algorithms used to determine how to fetch data over the network. Indeed, in the early years of streaming, the problem of starting fast while building the buffer was a key problem, solved at first by bursting the transmission rate [2], then by slowing down the playback [3], and finally by playing back at lower quality until the buffer is established [4]. For spherical video, however, there is user interactivity not just at the start. Any time during playback the user can quickly change where she is looking, with a momentary glance or turn of the head to change the azimuth and elevation of the viewport. Thus streaming for spherical video must be *user-adaptive*. To keep average bandwidth low while keeping the quality perceived by the user high, it is necessary to predict the variation in user behavior and pre-fetch the data in advance or else react quickly to make up any shortfall with low latency. In this sense, user-adaptivity is exactly analogous to network-adaptivity.

Streaming volumetric media, or holograms, for AR or VR must handle even higher levels of user interaction than streaming spherical video for VR. Since holograms support full 6DOF, or free-viewpoint, not only may a user turn her head to change her view direction, but she may also navigate freely among a multitude of holograms, changing both her view direction and position in space, in the process potentially changing her proximity to the various holograms over a wide range of distances, and changing the direction from which she sees the holograms.

In this paper, we address the problems of streaming holograms in AR/VR, using novel approaches to accommodate the higher level of user interactivity. We introduce the notion of a *window* as a buffer, replacing the traditional notion of a queue as a buffer. Using a window rather than a queue, it is possible to respond quickly to an unforeseen user action by inserting updated content just-in-time before it is played back, rather than inserting it at the end of the queue and subjecting the user to large latencies. When network behavior and user behavior are predictable, the window acts as a queue. We also extend the notion of 2D tiles used in spherical video to 3D tiles for holograms, so that different tiles may have different resolutions depending on where the user is positioned and looking relative to each tile, thus saving bandwidth by focusing on what the user is looking at. While the concept of 3D tiles may seem straightforward, there are several complicating issues. First, many of the 3D tiles, as regions of space, are empty a significant part of the time. Thus, even if the user viewpoint is not changing, the set of tiles occupied by content and visible to the user is content-dependent and changing. Especially considering that there can be many more tiles in 3D than in 2D, this requires an efficient way to index and address the tiles for holograms in AR/VR not needed for spherical video in VR. Second, in 3D the tiles may be occluded or far away. When streaming spherical video, the distance from the user to the scene is not variable. However, when streaming holograms, the scale of a tile in the user's view, and hence the utility of its decoded representation, can change dramatically as an inverse function of the user's distance to the tile, and may also be limited by the resolution of the rendering device. We develop a model of the utility of the tile representations based not only on its bitrate and whether the tile is visible, but also on the distance of the tile from the user. Finally, we develop a greedy yet optimal algorithm for maximizing the utility of the tiles subject to a rate constraint and couch it within the client buffer manager of the streaming client, to request an optimal set of tiles at every transmission opportunity. The algorithm naturally handles bitrate allocation between multiple holograms streaming simultaneously, as well as their playback at different speeds. We believe our work is the first to address streaming of volumetric media or holograms for AR or VR.

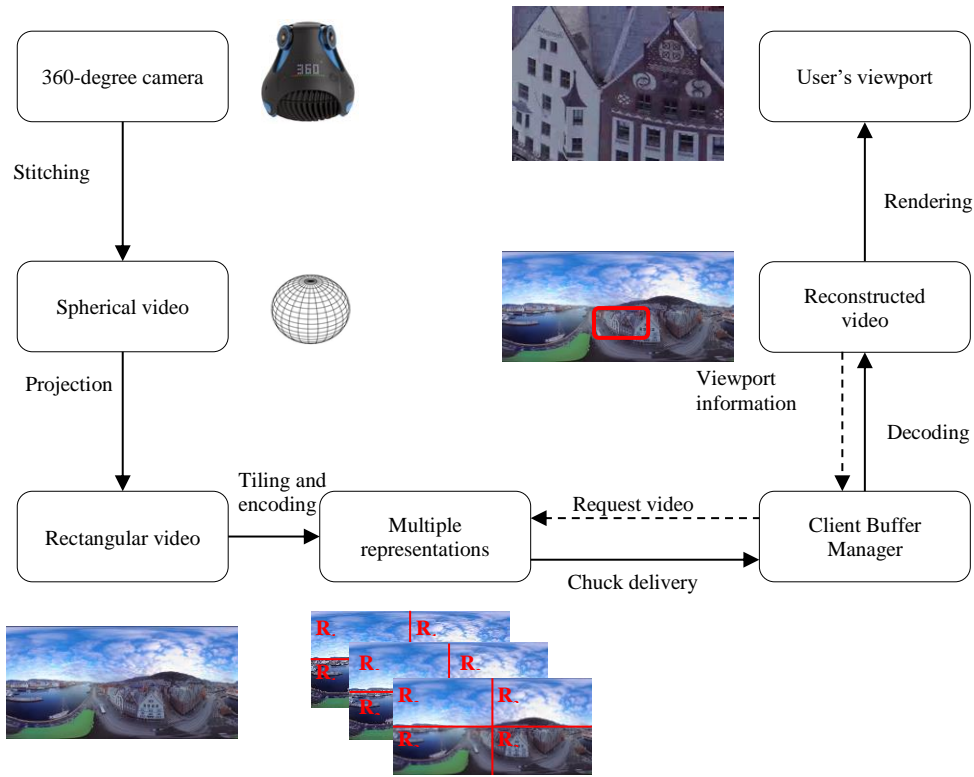

Fig 2. Tiled spherical video system

An overview of our system for streaming volumetric media is shown in Fig. 1. On the left, a web server stores the volumetric media objects, or holograms, for streaming. Each object is represented as a sequence of voxelized point cloud (VPC) frames, the frames are grouped into groups of frames (GOFs), and the sequence of GOFs is divided temporally into segments. Each segment is independently compressed to a small set of representations, each representation at a different bitrate. The bitrates are constant across all segments. Each GOF is divided spatially into tiles, which are independently coded.

A media presentation description (MPD), or manifest, accompanies each object to describe the collection of representations, while a segment index accompanies each segment to index the collection of tiles within the segment. On the right, the buffer manager in an HTTP client downloads the manifest, downloads the segment index for each segment in the window, estimates the throughput, estimates the user position, calculates the optimal collection of 3D tiles that will maximize utility for a given rate constraint, requests the tiles, downloads them into the buffer, advances the window, releases the tiles that fall out of the window for rendering, decoding, and presentation by the application, and repeats.

Experimental results show, regarding network adaptivity, that our system has higher throughput and robustness compared to previous state of the art algorithms. Further, regarding user adaptivity, it offers lower latency and higher quality than alternative designs, because the buffer is a window rather than a queue, and our utility maximization algorithm knows how to manage such a window.

This paper is organized as follows. Section II covers related work. Section III details our system. Sections IV and V detail our utility functions and rate-utility optimization algorithm. Section VI presents experimental results, and Section VII concludes.

## II. RELATED WORKS

A. Network-adaptive video streaming

Throughput and latency through the Internet from a streaming server to a client is not always stable. Therefore, the content must be buffered by the client to ensure continuous playback. During periods where the network throughput momentarily drops or even goes to zero, the content already in the buffer may be played back without the user noticing a stall, or *rebuffering* event. Furthermore, the content quality (as a function of its bitrate) may have to be varied over time to adapt to changing network conditions. During periods where the network throughput drops for an extended period, the quality of the content may have to be reduced to match the lower throughput. If the network throughput increases later, then the quality of the content may be increased, or alternatively, the duration of the buffer may be increased, or both. How to adapt the bitrate of the content as a function of what is happening with the network and the state of the buffer, while avoiding stalls and giving the user the maximum average quality without too many quality changes, is known as *client buffer management*, and has been the core issue in streaming media to the present time.

The early work on video streaming assumed packet-level communication between a server and client. Practical systems were frequently built on top of lossy protocols such as RTP/UDP and RTCP/UDP. The emphasis was on sender-driven streaming, although client-driven approaches were considered. As an example, Chou and Miao [5] performed rate-distortion optimized streaming of packetized media in an environment where individual packets were transmitted or requested, whose delivery could be confirmed only after a round trip time, during which many packets would be "in the air" necessitating a Markov decision process model for determining which packets to transmit or request. The emphasis was on a continuum of quality enabled by fine grained scalable (FGS) coding.

Eventually the early lossy, packet-level, FGS streaming was simplified by chunking the packets into longer segments of time, transmitting them by a reliable protocol (TCP), and chunking the quality into a sparse set of multiple bitrate (MBR) representations, which were simulcast. [6][7] was one of the first works in this area, proposing a receiver-driven approach to streaming of MBR content. By transmitting the chunks over HTTP/TCP as the reliable protocol, this receiver-driven approach became ideally suited to clients of web servers, and thus became the basis for practical systems and protocols such as Microsoft Smooth Streaming, Adobe HTTP Dynamic Streaming (HDS), Apple HTTP Live Streaming (HLS). The approach was eventually standardized as Dynamic Adaptive Streaming over HTTP (DASH) [8]. In these approaches, the streaming client appears to the web server as a web client, issuing HTTP GET requests to obtain the data. (In practice, a content distribution network, or CDN, intercepts such requests and returns the requested data from a nearby cache. Whether the data is returned from the original web server or from a CDN is not directly visible to the client.) The client first obtains a manifest, known in DASH terminology as the media presentation description (MPD), which describes the media chunks, e.g., the segments and their bitrates, for a stream. The client buffer management algorithm then uses the manifest to determine which chunks to request, and when to request them, depending on how long it takes for the requested chunks to arrive over the network.

The DASH protocol does not specify which client buffer management algorithm to use. An implementor is free to implement any algorithm. Various algorithms have been proposed [9]. Recent client buffer management algorithms model the client buffer as a queue of segments. Each segment can have a different bitrate. The client buffer manager (CBM) simply determines the bitrate for the next segment, requests from the server the chunk with that bitrate for the next segment, receives the chunk, and puts it at the end of the queue. Thus, the only algorithmic question is which bitrate to request for the next segment, and when to request it.

Recent client buffer management algorithms fall into two categories: throughput-based and buffer-based algorithms. With throughput-based algorithms, clients request the video bitrate for each segment based on the client's estimate of the current network throughput. This is an intuitive way to select the video bitrate. However, throughput estimation error can cause buffer depletion, leading to rebuffering events. Jiang et al. [10] introduced a throughput-based algorithm, FESTIVE, that improves throughput estimation by considering outliers. Luca el al. [11] introduced a throughput-based algorithm, ELASTIC, that eliminates the on-off transmission pattern that makes throughput estimation difficult, thereby making the estimation more robust. Li et al. [12] reduced the throughput estimation error by taking the on-off pattern into account. These algorithms improve throughput estimation accuracy, but do not consider buffer status to achieve better performance.

With buffer-based algorithms, clients request the video bitrate for each segment based on the buffer occupancy. When the buffer occupancy is high, which means that the buffer has enough video frames to play without receiving new frames for quite a while, a client may request a video segment with a high bitrate even though it may take a long time to download. Otherwise, the client may request a low bitrate, in order to download the segment quickly and build the duration of the buffer. Huang et al. [13] proposed a buffer-based algorithm that considers only the current buffer occupancy to request the bitrate of the next segment. Used by Netflix for long entertainment videos, in steady state it typically builds over 90 seconds of buffer occupancy, making playback in the absence of much interaction extremely robust. Spiteri et al. [14] provide a non-linear model to select better video bitrates among a limited number of choices.

Unfortunately, neither rate-based nor buffer-based algorithms are well suited for VR or AR streaming applications, in which user interaction is frequent. Buffer-based algorithms rely on large buffer durations for their stability, while rate-based algorithms pay little or no attention to buffer duration. Neither is appropriate when user interactions are frequent, because the buffer duration translates directly to the delay experienced by a user between an interaction by the user (say a head turn) and a change in the content being played back. Such a delay should be sub-second, certainly not much longer.

B. Tile-based streaming of spherical video

Streaming spherical video is more challenging than streaming ordinary video, because spherical video has orders of magnitude more pixels for the same angular resolution as ordinary video. However, a user can foveate on only one portion of the spherical video at any given moment. Thus, large portions of the spherical video behind the user or away from the foveal region may be streamed with much lower bitrate than the portion in the foveal region [15]-[20] . One way to achieve this is with tiles [15][16][18][20]. In tile-based streaming, the spherical video is partitioned spatially into smaller tiles, which are independently encoded and packaged as video streams at different bitrates. The client buffer manager tracks not only the network throughput and/or buffer status, but also tracks the user's view direction, and requests different representations for each tile depending on the user's view direction, and more importantly, depending on the client buffer manager's prediction of what the user's view direction will be at the time the requested representations will be played back to the user after making their way through the buffer. Fig. 2 shows the architecture of a tiled spherical video streaming system.

Since multiple tiles are delivered at the same time through the same network resource, which has limited bandwidth, the client buffer manager must allocate the available bit rate among the tiles, to provide the best experience to the client. Alface, Macq, and Verzijp [15] introduced an algorithm that selects a bitrate for each tile to maximize a utility function subject to an overall bitrate constraint. The utility function is based on the probabilities that the user will be looking at different tiles, at the time that the tiles are played back to the user, $T$ seconds in the future if $T$ is the current buffer duration. The algorithm selects higher bitrates for tiles with higher probability to be looked at $T$ seconds in the future. Because prediction of the user's view is imperfect, even when $T$ is only a few seconds, most tile-based bitrate selection algorithms elect to spend at least a minimal bitrate on every tile, so that the user will see at least a minimum level of quality even with an unexpected head turn.

C. Volumetric media communication

Volumetric media have existed at least since the invention of the Gabor hologram in 1948 [21]. In recent years they have been created using depth cameras and surface reconstruction, e.g., [22][23][24], rather than laser interferometry.

To date, communication of volumetric media has been driven mostly by immersive telepresence, e.g., [25][26][27][28][29][30]. Immersive telepresence, now sometimes referred to as holographic telepresence, holopresence, or holoportation, is two-way interactive communication like telephony, which traditionally involves very different communication protocols and algorithms than streaming media on demand or live broadcast, principally because the latency and fan-out requirements are so different.

To our knowledge, there has been no published work to date on streaming volumetric media on demand or live broadcast. One work comes close, in that it addresses the creation of "streamable free-viewpoint video" [31]. In [31], a high-quality point cloud is computed for each frame from dozens of RGB and IR cameras in an outside-in configuration. The point cloud is used to create a watertight mesh using Poisson surface reconstruction, and surface colors are computed by blending information from the RGB cameras. Keyframes are selected and the group of frames (GOF) between keyframes is re-meshed into a temporally-consistent dynamic mesh. The sequence of color texture maps within each GOF is compressed as a group of pictures (GOP) in an H264 video sequence inside an MP4 file, and the associated geometry is compressed separately and included in the MP4 file in a custom Network Abstraction Layer (NAL) unit, suitable for streaming. However, no streaming experiments are performed, and no protocols or algorithms that may address the increased user interactivity requirements of volumetric media streaming for applications such as AR are discussed.

Compression of volumetric media is currently in a period of rapid improvement. Early work assumed that volumetric media were represented as meshes. Therefore, mesh compression techniques such as geometry images/video [32][33], and graph wavelets [34][35] were examined. However, as point cloud capture became more prevalent, later work dropped the requirement of compressing meshes per se and turned towards compressing only what is necessary to reconstruct the geometry and its color attributes at the decoder. Thus volumetric media began to be represented and compressed as point clouds [36]-[46]. Point cloud compression (PCC) is now the subject of an MPEG standardization effort [47]. The point cloud codecs generally assume the point clouds are *voxelized*. That means the points $(x, y, z)$ are quantized to a voxel grid within a *bounding cube*. A voxel is said to be *occupied* if it contains a point, and any $N \times N \times N$ grouping of voxels is said to be *occupied* if it contains any occupied voxels. By grouping voxels together into independently coded tiles, it is possible to extend the basic idea of tile-based streaming from spherical video to volumetric media. However, unlike the tiles in spherical video, the set of occupied tiles is sparse, changing, content-dependent, and possibly large. Furthermore, the set of visible tiles is not only content-dependent but also viewpoint-dependent due to occlusions.

How to stream such tile-based partitions of volumetric media in the presence of user interactivity is the subject of our work.

III. WINDOW-BASED CLIENT BUFFER MANAGER

In this section, we detail several concepts of our window-based client buffer manager (CBM): the window-based buffer, the file format and how it supports multiple objects and tiles, the interface between the client buffer manager and the playback engine to support user interaction, and the CBM operation.

A. Window-based buffer

When a user wearing a head mounted display turns her head, or picks up an object to examine it, a portion of the content, which until that moment may have been streaming at low resolution, is immediately required to be displayed at high resolution. The same thing can happen with a handheld display, for example when an object is streamed to a mobile phone, and the user suddenly spins the object around or zooms in to see a portion of the object at higher resolution.

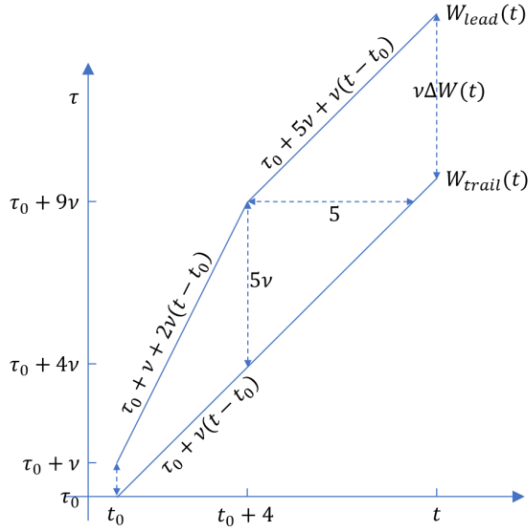

Fig 3. Evolution of window over time

In traditional streaming, the buffer is a queue. The length of the queue in terms of seconds of content in the buffer is typically on the order of five seconds (or over 90 seconds in the case of a typical buffer-based algorithm), to make the streaming robust to underflow in the event of an unforeseen drop in network throughput. Unfortunately, a long queue means that the user experiences a long delay between her action (e.g., head turn) and a change in the level of fidelity or resolution needed in response. This is unacceptable in AR and VR applications, where user interaction is frequent.

Our solution is to eliminate the notion of the buffer as a queue. Instead, we treat the buffer as an interval, or *window*, that advances over the media presentation timeline. The leading edge of the window corresponds to the tail of the traditional queue, where groups of frames (GOFs) are traditionally inserted, while the trailing edge of the window corresponds to the head of the traditional queue, where GOFs are traditionally extracted for decoding and playback. However, a window-based CBM may ask the server for any GOFs inside the window, not just a GOF at the leading edge of the window. Moreover, a window-based CBM may ask for a GOF in the window that has already been transmitted and may be sitting in the buffer at a low resolution. For example, the CBM may ask for a higher resolution version of a GOF that is about to exit the trailing edge of the window and be played back to the user, even if it already exists in the buffer at a lower resolution. In this way, a window-based CBM is able to respond with low latency to unexpected user interactions.

Fig. 3 shows the evolution of a window over time. The horizontal axis is user time $t$, for example, the user's wall clock time. The vertical axis is media time $\tau$, for example, the wall clock time at which the content was recorded. The figure shows two monotonically increasing functions: $W_{trail}(t)$ and $W_{lead}(t)$. At any given user time $t$, the vertical interval $[W_{trail}(t), W_{lead}(t)]$ between these two functions is the window of content that may be in the client buffer at any given user time. As user time $t$ advances, the window moves upward. Content leaving the trailing edge of the window is released to the decoder for playback. Content entering the leading edge of the window is newly available for the client to request from the server.

Specifically, if the user seeks to media time $\tau_0$ and plays the content at speed $v$ beginning at time $t_0$, then at any time $t \geq t_0$ the content at media time

$$W_{trail}(t) = \tau_0 + v(t - t_0)$$

is released from the buffer as it leaves the trailing edge of the window, so that it can be decoded and played back to the user. At the same time $t$, the client buffer manager may fetch content as far ahead as media time

$$W_{lead}(t) = W_{trail}(t) + v\Delta W(t - t_0),$$

where $\Delta W(t)$ is a window size function. The window size function is the number of seconds of user time in the buffer. The window size function that we use in our experiments, illustrated in Fig. 3, is

$$\Delta W(t) = \begin{cases} 1 + t & 0 \leq t \leq 4 \\ 5 & 4 \leq t. \end{cases} \quad (1)$$

For this function, the window size grows from 1 to 5 seconds over the first 4 seconds of playback time, then remains at 5 seconds. The purpose is for window size to start small to allow playback to begin quickly, but to grow soon to allow playback to be robust to network dropouts. Other functions besides (1) are also possible, such as the logarithmic function $\alpha + \beta \log(t)$ used in [5].

Early work (by one of us) used this kind of a window-based approach for rate-distortion optimized streaming of packetized media [5]. In that work, in the case of client-driven streaming, the client used a rate-distortion criterion to decide which packet within the window to request at each request opportunity. Because a packet is small, the time to transmit it is generally small in comparison to the round trip time. Moreover, the packet could be lost or delayed, creating uncertainty about whether it would be received by the decoding deadline, and therefore uncertainty about whether other packets that depend on it would have any value. These factors forced [5] to use a relatively complex Markov decision process to make its rate-distortion decisions. In the present paper, we take an analogous window-based approach to rate-utility optimized streaming, but instead of requesting a single packet at a time, our CBM requests large chunks of data at a time from a web server. The return of the data can be modeled as happening immediately, because the round trip time is negligible compared to the transmission time. Furthermore, the requested data can be modeled as returning with certainty, because the data are transmitted over a reliable protocol (HTTP/TCP). This greatly simplifies the decision logic in the CBM. The rate-utility optimization algorithm that we use in this paper is optimal, but it reduces to a simple greedy algorithm, as explained in Section V.

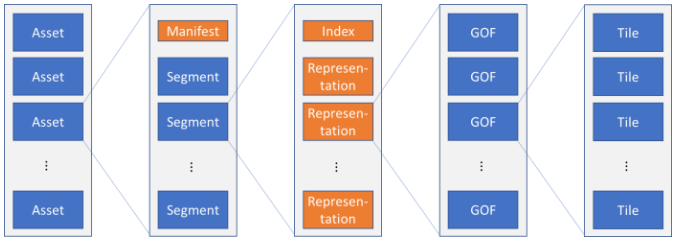

Fig 4. Structure of files available on server. Individual files are shown in orange

B. Multiple objects

In streaming holograms, unlike streaming video or even streaming spherical video, it may be common to stream multiple objects simultaneously. Each object represents a single performance, such as a single person performing a single pose. In general, multiple objects will be streamed together and composited at the client. The client must know how to composite the objects by spatio-temporal transformations, which are specified elsewhere. The objects may come from different web servers.

In our framework, each of the objects that play back simultaneously may have a different seek time $\tau_0$ and playback speed $v$. However, all such objects are assumed to begin playback at a common user time $t_0$ and have a common window function $\Delta W(t)$. Thus the windows onto each object may have different media times, but they coincide in user time.

C. File format

Here we outline a possible file structure consistent with DASH, with some extensions or modifications (noted in italics) to handle tiled holograms. Other file structures are also possible.

Fig. 4 shows the structure of files available on the server. There are three types of files: *manifests*, *segments*, and *segment indices*. The manifest for an object is the meta data for the object, as it contains a list of all the other files for the object, and their properties. The segments contain the data of the object, as they are the elements of a temporal partition of the object's media into intervals of equal duration. Each segment comes in various *representations*, which are different qualities, spatial resolutions, or temporal resolutions. A segment index accompanies each segment, to index into representations, GOFs, and tiles of the segment. These file types are now described in detail.

*Manifest*

The manifest for an object comprises the object's properties, a list of its representations, and a schema or *template* for mapping a desired time segment and representation to a URL, for example:

```
Type (static=on demand, dynamic=live)
ProgramInformation
  title
Period[periodCount]
  duration
  AdaptationSet[adaptationSetCount]
    maxwidth
    maxFrameRate
    cubeToObjectScale
    cubeToObjectRotation
    cubeToObjectTranslation
    tileWidth
    startTime
    mimeType
    codecs
    SegmentTemplate
      timescale
      duration
      media
      startNumber
    Representation[representationCount]
      id
      bandwidth
      width
      frameRate
```

`Period` contains all content within a certain `duration`. There may be multiple `Period`s.

`AdaptationSet` contains content that can be selected by the user (or programmatically by the user's agent) to be played during the period. In standard DASH, each `AdaptationSet` is typically a video, audio language 1, or audio language 2. When extended to volumetric media, each `AdaptationSet` may contain a single dynamic hologram (i.e., an object) or its audio track. An `AdaptationSet` is either selected or not selected in its entirety. If the `AdaptationSet` is a hologram, `maxwidth` is the width of the bounding cube of the hologram in voxels, `maxFrameRate` is the maximum frame rate of the hologram, `cubeToObject` is the transformation from the bounding cube coordinate system into the object's coordinate system, `tileWidth` is the width of each tile in voxels (relative to `maxwidth`), `startTime` is the media start time ($\tau_0$) within the period, `mimeType` is a new MIME type/subtype such as *holo/hvr*, and `codecs` is the name of the codec.

`SegmentTemplate` is the template for all the segments in the `AdaptationSet`. `Duration` is the length of each segment (in `timescale` units). `Media` is the template in a text format such as

`ObjectName_$bandwidth$_$width$_$framerate$_$number$.hvr`

`StartNumber` is the `$number$` of the first segment, typically 0. `$bandwidth$`, `$width$`, and `$framerate$` indicate the representation used.

`Representation` contains information about the representations available for the `AdaptationSet`. The information includes `id` (a text description), `bandwidth`,

`width` (i.e., the number of voxels across the bounding cube, in this representation), and `framerate` to indicate the bandwidth and level of detail (LOD) of the representation.

*Segment*

The segment `ObjectName_$bandwidth$_$width$_$framerate$_$number$.hvr` is the chunk of the media file containing an encoding of the object from time ($number$ - startnumber) * duration up to but not including time ($number$ - startnumber + 1) * duration, in the representation with the given `$bandwidth$`, `$width$`, and `$framerate$`. An initial segment can also be specified.

A segment consists of a sequence of GOFs (groups of frames). Each GOF consists of an optional GOF header followed by a sequence of tiles. Tiles that contain content are *occupied*. Only occupied tiles appear in each GOF. The tiles appear in Morton order, and are indexed by their Morton codes. The Morton code of a tile is the interleaving of the bits for the coefficients of the tile's position $x$, $y$, and $z$ in the bounding cube coordinate system.

*Segment Index*

There is one segment index file for each segment. The segment index is an index into its associated segment and all of its representations. The segment index will be used by the client to identify particular representations for particular tiles in particular GOFs in the segment. The segment index needs to be compact, and (like the segment) is a binary file. If a segment is stored in a file

`ObjectName_$bandwidth$_$width$_$framerate$_$number$.hvr`

then the segment index can be stored in an associated file

`ObjectName_$number$.idx`

(or in another file designated by a filename extension decided a priori).

The segment index contains pointers to the GOFs and to the tiles of the segment, so that the tiles for each GOF can be downloaded independently as desired. The segment index information includes:

```
gofCount
Gof[gofCount]
 startTime
 duration
 frameCount
 tileCount
 Tile[tileCount]
  mortonCode
  normalCode
 Representation[representationCount]
  gofByteOffsetInSegment
  gofHeaderByteCount
  Tile[tileCount]
   byteCount
```

`StartTime` is the start time of the GOF; `duration` is its duration. `FrameCount` is the number of frames in the GOF. `TileCount` is the number of tiles in the GOF. `MortonCode` is the Morton code for a tile. `NormalCode` is a code for the dominant normal direction of the tile. Both position and normal direction may be used to test whether a tile is visible. For each representation of the segment, `GofByteOffsetInSegment` is the byte offset of the GOF in the file for its segment and representation, `GofHeaderByteCount` is its size, and `Tile[n].byteCount` is the size of the nth tile. Note that all representations for a segment have the same GOF and tile structure. Only byte offsets and counts may be different across the segment index files for different representations of the same segment.

If visibility testing is not required, there need be only a single tile (the whole cube, having Morton code 0). If visibility testing is desired, there may be hundreds of tiles per GOF, depending on the granularity of the visibility testing. Even with a hundred tiles, four representations, 32 bits per `mortonCode` and `byteoffset`, and GOF size of 4 frames, the bitrate for the segment index should be at most only about 100*(32+4*32)*30/4 = 120 Kbps, which is about 1% of a typical total bit rate. However, we may wish to have many frames per GOF, few tiles per GOF, and few representations to keep the segment index bitrate as low as possible. Or the client can fetch only the relevant parts of the segment index (e.g., for low-bitrate representations).

When subsets of nearby tiles are accessed, they can be grouped together into a single byte range, as part of a multi-part byte range request for the tiles in the segment. This is made efficient by having the tiles in Morton order.

D. Client buffer manager interfaces

The client buffer manager (CBM) has two interfaces: one to the web server or CDN and one to the playback engine (PE).

The interface between the CBM and the web server is solely through HTTP requests, for efficiency preferably HTTP/2 [48], which allows the client to communicate with the server in a binary format to create multiple streams that can be prioritized and for dependencies between streams all over a single connection.

The interface between the CBM and the PE could be something like the following.

- The PE tells the CBM the URLs of the manifests of the objects that it wishes to stream, the seek time, the playback speed, and whether to loop or not. It also provides initial user point(s) of view in world coordinates (i.e., 3D user position, azimuth, elevation, and horizontal FOV). Finally, it provides some device resolution information: pixels across the horizontal FOV. The latter is used to evaluate the utility of different levels of detail.

- The CBM opens the manifests, downloads the initial segment and the segment index of the first segment(s) of each representation of each object, downloads the lowest bitrate representation of the objects for a default period of media time (e.g., $v$ seconds), places the received representations in the buffer, and notifies the PE that at least $v$ seconds of content is in the buffer so that it can begin playback.

- The PE notifies the CBM every time it pulls something out of the buffer, so that the CBM can update the head of the buffer.

- The PE also notifies the CBM of any change in the user frustum or frusta, which will be any time the user updates his or her position or look direction relative to the object.

- The CBM will continue to download content into the buffer. It will notify the PE on any significant change to the buffer. For example, if the PE tries to access the buffer and it is empty (but not yet EOF), then the buffer underflows. The CBM will notify the PE when there is once again $v$ seconds in the buffer to begin playback.

E. Client buffer manager operation

At the core of the operation of the CBM is a data structure for the buffer that mirrors the structure of the files on the server. There is an array of objects. Each object contains information from the manifest, and an array of segments. Each segment contains an array of GOFs. Each GOF contains an array of (occupied) tiles. Thus in this data structure a tile is not only spatial but also temporal (as an element of a GOF in a segment) and is furthermore associated with an object. Each tile (in each object, segment, and GOF) contains the *utility*[$m$] and *bitCount*[$m$] for each representation $m = 1, ..., M$ of the tile, and contains the selected representation $n$ of the tile. The initial representation $n$ is set to 0, meaning that the tile initially has no representation, and by convention *utility*[0] = *bitCount*[0] = 0. Once the representation for a tile is actually fetched, then the tile also contains the fetched data. The CBM may request a newer (better) representation for a tile at any time, in which case the older representation of the tile is overwritten.

The representation of a tile at the current playback time is fed to the PE to decode and render. Although not illustrated in the pseudo code in the Appendix, it is possible for the PE to loop playback, in which case the CBM will continue to upgrade the representations on every subsequent pass until there are no more upgrades possible.

The CBM thread begins when the CBM is notified by the PE to begin playback. The, the CBM downloads the manifests of the specified objects, determines the initial segments, downloads the indexes for those initial segments, determines the GOFs required to play the first $v$ seconds of each object, requests the lowest bitrate representation for all the tiles in those GOFs, and waits to receive the requested data from the server. When the data are received, the start time $t_0$ is declared, and the CBM notifies the PE to begin playback. From that point on, the "buffer" at time $t \geq t_0$ is the set of all tiles for which data has been received (in all GOFs in all segments in all objects) whose media times are within the window (shown in Fig. 3) at time $t$. Furthermore, from that point on, the PE on its own thread extracts data from the buffer at the trailing edge of the window and decodes it for presentation.

Beginning at time $t_0$, the CBM enters a request-wait cycle, with the CBM requesting data from the server at each *request opportunity* $t_i$, $i = 0, 1, 2, ...$, and between requests, waiting for the data to be returned. At each request, the CBM requests just enough data so that the expected length of time for the data to be transmitted from the server to the client (the "on" part of the cycle) is $T$ seconds. $T$ is chosen large enough so that the length of time for the request to go out to the server (the "off" part of the cycle) is a negligible fraction of the cycle, so that that the downlink bandwidth is nearly fully utilized. At the same time, $T$ should be small enough so that the client can quickly adapt to unexpected changes in the network or user behavior. We use $T = 0.5$ seconds in our experiments.

From the CBM point of view, all the action happens at the request opportunities. Between request opportunities, the CBM simply waits. At request opportunity $t_i$, the CBM first estimates the network throughput $C_i$ by smoothing the instantaneous estimate $\hat{C}_i = R_{i-1}/(t_i - t_{i-1})$, where $R_{i-1}$ is the number of bits requested at the previous request opportunity. A typical way to smooth the instantaneous throughput estimates is with a first-order autoregressive filter, $C_i = wC_{i-1} + (1-w)\hat{C}_i$, for some weight depending on the memory of the network. We use $w = 0.75$ in our experiments. This gives a bit budget of $R_i = C_i T$ for the current request.

Given the bit budget $R_i$ for the request, which tiles (within the window) and their representations to request are decided based on maximizing an expected "utility" to the user, subject to the bit budget. Specifically, if $\mathcal{W}$ is the list of tiles in the window, $\mathcal{N} = \{tile.n\}$ is the list of current representations for the tiles in $\mathcal{W}$, and $\mathcal{M} = \{tile.m\}$ is the list of representations for the tiles in $\mathcal{W}$ after the request is fulfilled, the CBM chooses $\mathcal{M}$ to maximize the total utility

$$U(\mathcal{M}) = \sum_{tile \in \mathcal{W}} tile.utility[tile.m] \quad (2)$$

subject to the constraint

$$R(\mathcal{M}) = \sum_{\substack{tile \in \mathcal{W} \text{ and} \\ tile.m \neq tile.n}} tile.bitCount[tile.m] \leq R_i. \quad (3)$$

Note that the latter sum is only over tiles in the window whose future representations *tile.m* differ from the current representation *tile.n*, since only then will the representation *tile.m* have to be requested from the server and its size in bits count towards the bit budget. A request for a new representation is made, if possible, by combining the requests for all tiles in the same segment with that representation into one multipart byte range request.

At every request opportunity, the trailing edge of the window (i.e., the current playback time) advances by some amount (namely the time that has elapsed since the last request opportunity), and the leading edge of the window also advances by the same amount (or a greater amount, at the beginning of playback, when the window size is still growing). Most of the requested tiles will fall into the part of the window that is newly opened. However, tiles may still be requested in other parts of the window, especially if there is a surge of new bit rate available, of if the user viewpoint moves and causes new tiles to become visible. The CBM is able to get these newly visible tiles within $T$ seconds, as it is not restricted to requesting tiles only at the leading edge of the window.

The utility of a tile is a function of (among other things) the user's view pose and frustum, which are passed from the PE to the CBM and must be kept updated as the user moves. If the user suddenly move her head, the utility of some tiles may increase, causing them to get higher priority and hence causing the CBM to request them in a higher quality or resolution than they received than at any previous request opportunity. How the utility is calculated is detailed in Section IV, and how the maximization is performed is detailed in Section V.

## IV. UTILTY MEASURES

The utility of a tile depends on how much useful information it brings to the user. This depends in turn on the relationship of the tile to the viewpoint of the user, as illustrated in Fig. 5. If a tile is outside the viewpoint of the user, then it has no utility. If it far away, then its utility may be low because it covers a small area. On the other hand, as the tile moves quite close, then its utility may saturate because its spatial resolution is intrinsically finite. And of course a tile coded with a higher bitrate will generally have a higher spatial resolution, higher temporal resolution, and/or higher PSNR. Unfortunately, the utility of a tile in the far future is uncertain, because the user's viewpoint in the future is uncertain. Hence in principle one can evaluate only the *expected* utility of a future tile. To complicate matters, frequently there are multiple points of view, certainly in stereoscopic systems, but possibly also in multi-user systems.

We model the (expected) utility of a tile as a function of the bandwidth of the representation selected for the tile, weighted both by the number of distinguishable voxels in the tile and by the probability that the tile will be visible (i.e., in the user's frustum and facing the user) when it is decoded and rendered.

To be more specific, we change notation from the previous section. For tile $k$, with representation $n_k$, and viewpoints $v \in \mathcal{V}$, the utility $U_k(n_k)$ of the tile is given by

$$U_k(n_k) = u(B_{n_k}) \times \max_{v \in \mathcal{V}}\{LOD_k(n_k, v) \times P_k(v)\},$$

where $u(B)$ is a function indicating the utility per distinguishable voxel of the tile as a function of the bandwidth of the representation, $LOD_k(n_k, v)$ is the number of distinguishable voxels of the tile if the tile is visible from the current viewpoint $v$, and $P_k(v)$ is the probability that the tile would visible when rendered, if the current viewpoint is $v$. The latter two factors depend on the spatial resolution of the representation, the size of the tile, the point of view relative to the tile, and the resolution of the viewing device. However, the only parameter that the CBM can control to maximize the utility is $n_k$. All three factors are described in detail next.

### A. Utility per distinguishable voxel

We model the utility of the tile per distinguishable voxel as an increasing function of the bandwidth of the tile's representation, as a proxy for the quality of the representation. Though the utility of quality to any given user is hard to quantify, in general it should be monotonically increasing, should flatten out at high bitrates, and should be zero when nothing is transmitted. Thus we model the utility as an affine function of the logarithm,

$$u(B) = \begin{cases} \alpha \log(\beta B) & B > 0 \\ 0 & B = 0 \end{cases}, \quad (2)$$

where $\alpha$ and $\beta$ are normalization coefficients that bring $u$ into the range $[0,1]$ for all bandwidths $B_m$ of the representations $m = 1, \dots, M$. These coefficients are then kept constant for the duration of playback. Many Mean Opinion Score (MOS) tests frequently show such a logarithmic pattern [49].

### B. Distinguishable voxels in a tile

We model the number of distinguishable voxels in a tile as the number of distinguishable voxels in the square area roughly covered by the tile, that is, the number of degrees of view linearly across the tile times the number of distinguishable voxels per degree of view linearly across the tile, squared. In turn, the number of degrees of view linearly across the tile is approximately the width of the tile divided by the distance of the tile from the viewpoint (times $180/\pi$). Furthermore, the number of distinguishable voxels per degree of view linearly across the tile is the minimum of the number of voxels per degree of view linearly across the tile and the number of pixels per degree of view across the display. This information is computed based on the user's view pose and frustum, and the display resolution, passed to the CBM from the PE.

To be specific, let the width of the tile in the real world be the width of the tile in voxels (object.tileWidth) times its cube-to-world scale (object.cubeToObjectScale), and let the field of view of the tile (in radians) at unit distance be approximated by its real-world width. Let position of the tile in the real world be its $(x, y, z)$ position in voxels (as determined from its Morton code) time its cube-to-world translation (object.cubeToObject plus object-to-world translations), and let the distance to the position of the tile from viewpoint $v$ be $dist(v)$. Then the approximate field of view across the tile in radians is

$$RAD_k(v) = \frac{object.tileWidth * object.cubeToObjectScale}{dist(v)}.$$

Next let the number of voxels in the tile per radian be the width in voxels of the bounding cube in representation $n$ (*object.representation[n].width*), divided by the width of the bounding cube in the real world ($object.maxWidth * object.cubeToObjectScale$), times the distance to the tile:

$$VPR_k(n, v) = \frac{object.representation[n].width * dist(v)}{object.maxWidth * object.cubeToObjectScale}.$$

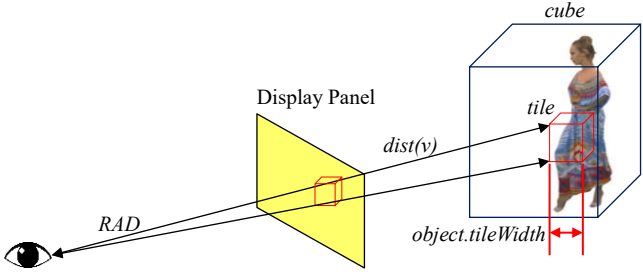

Fig 5. RAD

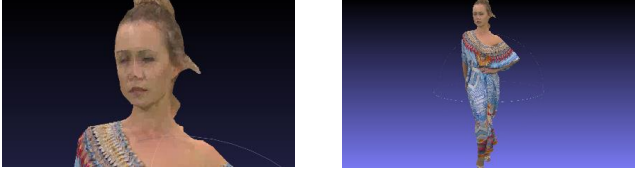

(a)  VPR<PPR  (b)  VPR>PPR

Fig 6. VPR and PPR

Finally let the number of pixels per radian across the display device be the number of pixels across the display (display.horzPixels) divided by the field of view of the frustum,

$$PPR_k(v) = \frac{display.horzPixels}{view[v].frustum.horzFOV}.$$

Then the minimum of $VPR_k(n,v)$ and $PPR_k(v)$ is the number of distinguishable voxels per degree across the tile, and

$$LOD_k(n,v) = [RAD_k(v) * min\{VPR_k(n,v), PPR_k(v)\}]^2 \quad (3)$$

is the number of distinguishable voxels in the square area roughly covered by the tile.

C. Probability that a tile will be visible

The last part of the utility model is the probability $P_k(v)$ that if the current viewpoint is $v$ then tile $k$ will be visible at by the time the tile emerges from the trailing edge of the window and is displayed to the user.

The uncertainty of whether tile $k$ will be visible by the time it emerges from the window is due, of course, to the uncertainty of the user's behavior in the interim. If the viewpoint of the user could be accurately predicted, then $P_k(v)$ could be set close to 0 or 1. If it were close to 0, then the utility of the tile would be close to 0, and no bits would need to be wasted transmitting the tile. The bits could be used instead to improve the quality of tiles for which $P_k(v)$ is close to 1.

Thus, for AR streaming, user prediction is an important problem, just as network prediction is important for all streaming. User adaptivity and network adaptivity are analogous. The problem of user prediction is therefore a new area for research in AR streaming.

In this paper, however, we keep the user prediction model very simple. Specifically, we predict that if a tile $k$ has media time $\tau_k$, it will be visible to the user when it emerges from the window if its position is visible to the user in the current view $v$ at time $t$, with prediction error probability 0.1 if the tile is early in the window (close to the trailing edge $W_{trail}(t)$), increasing linearly to 0.4 if the tile is late in the window (close to the leading edge $W_{lead}(t) = W_{trail}(t) + \Delta W(t)$). That is,

$$P_k(v) = \begin{cases} 1 - P_k^{err}(v) & \text{if } k \text{ currently visible from } v \\ P_k^{err}(v) & \text{otherwise} \end{cases} \quad (5)$$

Where $P_k^{err}(v) = 0.1 + 0.3 \min\{1, (W_{lead}(t) - \tau_k)/\Delta W(t)\}$.

This is a simple model of increasing uncertainty of what the user will be looking at further away in time. Undoubtedly the model can be easily improved using machine learning.

V.  UTILITY MAXIMIZATION

In this section, we present an algorithm for utility maximization that is greedy yet provably optimal. To begin, let us re-write the constrained maximization problem (2) and (3) as maximizing

$$U(\mathcal{M}) = \sum_{k=1}^{K} U_k(m_k)$$

subject to

$$R(\mathcal{M}) = \sum_{k=1}^{K} b_k(m_k) \leq R_i,$$

where $\mathcal{M} = \{m_1, \ldots, m_K\}$ are the representations for all tiles $1, \ldots, K$ in the window, and $b_k(m)$ is the number of bits that would be required to get representation $m$ for tile $k$. If $m = n_k$, the representation for tile $k$ that is already in the buffer, then $b_k(n_k) = 0$, because it takes no additional bits to get representation $n_k$ for tile $k$. If there is no representation for tile $k$ yet in the buffer, then $n_k = 0$ and still $b_k(0) = 0$. By convention, $U_k(0) = 0$.

Because we can restrict our search for the optimal $\mathcal{M}$ to the upper convex hull $\hat{\mathcal{S}}$ of the set of points $\mathcal{S} = \{(R(\mathcal{M}), U(\mathcal{M}))\}$ in the rate-utility plane[2], we can instead solve the easier problem of maximizing the Lagrangian

$$U(\mathcal{M}) - \lambda R(\mathcal{M}) = \sum_{k=1}^{K} [U_k(m_k) - \lambda b_k(m_k)]$$

for some $\lambda > 0$. Moreover,

---

[2] Not only does the convex hull $\hat{\mathcal{S}}$ outperforms $\mathcal{S}$ in the sense that for any point $(R, U) \in \mathcal{S}$, there exists a dominating point $(\hat{R}, \hat{U}) \in \hat{\mathcal{S}}$ such that $\hat{R} \leq R$ and $\hat{U} \geq U$, but also every point on $\hat{\mathcal{S}}$ can be achieved with timesharing or randomization between points in $\mathcal{S}$.

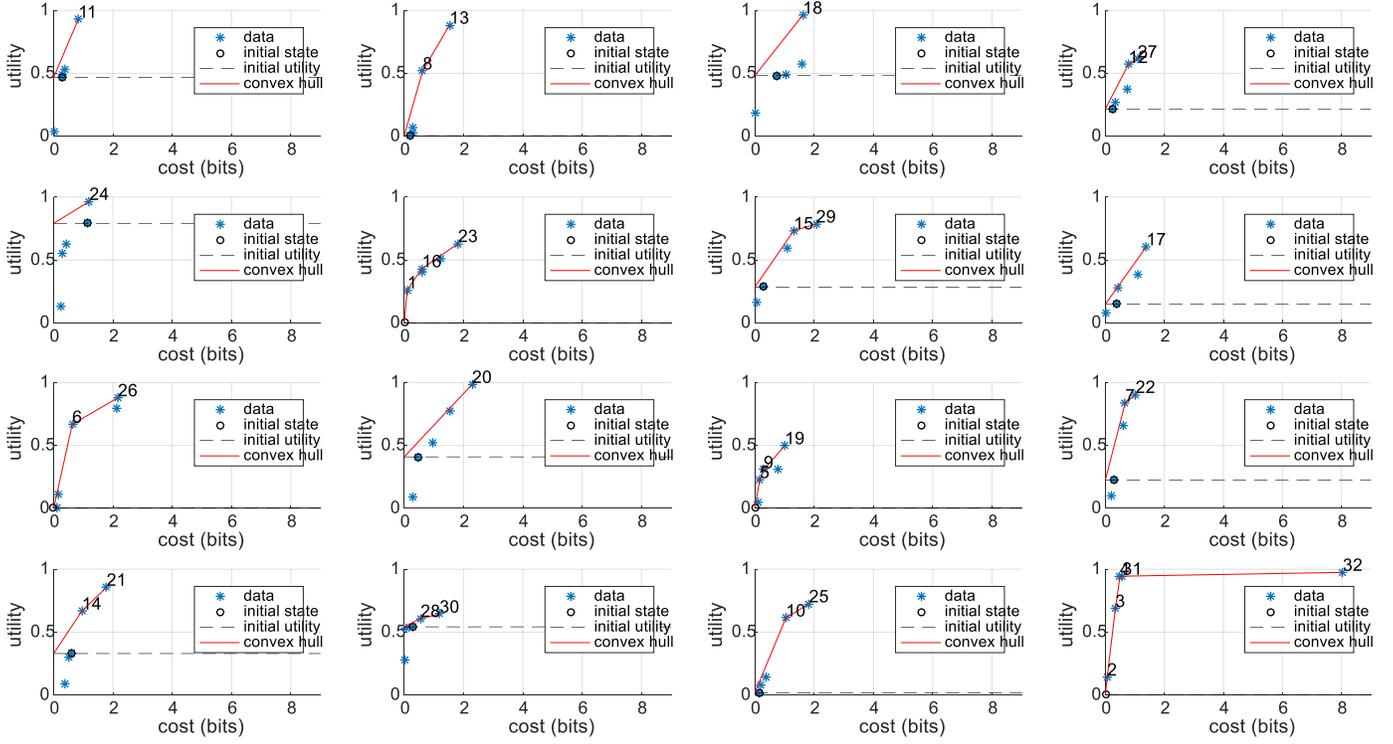

Fig 8. Rate-utility optimization for 16 tiles. Labeled numbers show order of selection. The algorithm chooses next point with highest slope, because it is next most efficient. Dashed lines indicate current status.

$$\max_{\mathcal{M}}[U(\mathcal{M}) - \lambda R(\mathcal{M})] = \max_{\{m_1,\ldots,m_K\}} \sum_{k=1}^{K}[U_k(m_k) - \lambda b_k(m_k)]$$

$$= \sum_{k=1}^{K} \max_{m}[U_k(m) - \lambda b_k(m)],$$

so the maximization problem can be solved independently for each tile. For each $\lambda$, the solution

$$m_k(\lambda) = \operatorname*{argmax}_{m}[U_k(m) - \lambda b_k(m)]$$

for tile $k$ lies on the upper convex hull $\hat{S}_k$ of the set of points $S_k = \{(b_k(m), U_k(m))\}$ in the rate-utility plane, and the points on the vertices of the convex hull are swept out in order of increasing $b_k(m)$ as $\lambda$ decreases from infinity to zero.

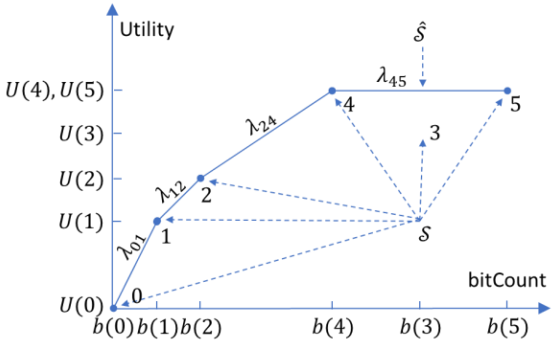

Fig 7. Rate-Utility points for a tile

As an example, Fig. 6 shows (with the tile index $k$ suppressed) a set of six rate-utility points $S =$ $\{(b(m), U(m)): m = 0, \ldots, 5\}$, with index 0 corresponding to the null representation and indices 1-5 corresponding to five representations for the tile's object in order of increasing *object.bandwidth[m]*. Points 0, 1, 2, 4, and 5 lie on the upper convex hull $\hat{S}$ in order of increasing $b(m)$. Let $\lambda_{01}$, $\lambda_{12}$, $\lambda_{24}$, and $\lambda_{45}$ be the slopes of the line segments between these points on $\hat{S}$. Then the optimal representation for this tile for any given $\lambda$ is

$$m(\lambda) = \begin{cases} 0 & \lambda_{01} < \lambda < \infty \\ 1 & \lambda_{12} < \lambda \leq \lambda_{01} \\ 2 & \lambda_{24} < \lambda \leq \lambda_{12} \\ 4 & \lambda_{45} < \lambda \leq \lambda_{24} \\ 5 & 0 \leq \lambda \leq \lambda_{45} \end{cases}.$$

For this tile, $\lambda_{01}$ is a threshold for $\lambda$ above which no representation is requested. The maximum such threshold across all tiles is a threshold for $\lambda$ above which no representations are requested for any tiles. As $\lambda$ decreases from this threshold, $b_k(m_k(\lambda))$ increases for every tile $k$, and hence $R(\mathcal{M})$ also increases. Thus $\lambda$ can be decreased step-by-step until the constraint $R(\mathcal{M}) \leq R_i$ would be violated. For this value of $\lambda$, the representation $m_k(\lambda)$ for tile $k$ can be requested from the server if $m_k(\lambda) > 0$. This is our basic rate-utility optimization algorithm.

One modification to the basic algorithm is crucial to be able to update, at a request opportunity $t_i$, the representation of a tile that remains in the window and already has a representation from request opportunity $t_{i-1}$, for example, if the tile suddenly increases in utility because the user has turned to look at it. In the modification, the representation of tile $k$ from previous request opportunities is saved in variable $n_k$, and $b_k(n_k)$ is set to 0, the rationale being that to obtain representation $n_k$ again at the current request opportunity would take 0 bits. The utility $U_k(n_k)$ is left unchanged. With this modification, the initial point on the upper convex hull for tile $k$ is $(0, U_k(n_k))$ rather than $(0,0)$. This raises the initial point of the upper convex hull, thus flattening the convex hull, making it difficult to reach other representations along the upper convex hull unless $\lambda$ is allowed to become large (e.g., if the estimated bit budget $R_i$ suddenly becomes large) or unless some other representation suddenly increases in utility (e.g., if the user turns her head to look at the tile). The full algorithm is shown in Table 1, and illustrated in Fig. 7.

TABLE 1.
RATE-UTILITY MAXIMIZATION ALGORITHM

Set $R_{current} = 0$.
For each tile $k$:
  Set $m_k = n_k$ (the existing representation)
  If $b_k(n_k) = \max_m b_k(m)$ then set
  $$\lambda_k^* = 0 \text{ and } m_k^* = n_k.$$
  Else set
  $$\lambda_k^* = \max_{m:b_k(m)>b_k(n_k)} \frac{U(m) - U(n_k)}{b_k(m) - b_k(n_k)},$$
  $$m_k^* = \underset{m:b_k(m)>b_k(n_k)}{\operatorname{argmax}} \frac{U(m) - U(n_k)}{b_k(m) - b_k(n_k)}.$$
while $R_{current} < R_i$
  Let $k = \max_k \lambda_k^*$ be the tile with the greatest $\lambda_k^*$.
  If $\lambda_k^* \leq 0$ then break.
  Update $R_{current} = R_{current} + b_k(m_k^*) - b_k(n_k)$.
  Set $n_k = m_k^*$.
  If $b_k(n_k) = \max_m b_k(m)$ then set
  $$\lambda_k^* = 0 \text{ and } m_k^* = n_k.$$
  Else set
  $$\lambda_k^* = \max_{m:b_k(m)>b_k(n_k)} \frac{U(m) - U(n_k)}{b_k(m) - b_k(n_k)},$$
  $$m_k^* = \underset{m:b_k(m)>b_k(n_k)}{\operatorname{argmax}} \frac{U(m) - U(n_k)}{b_k(m) - b_k(n_k)}.$$
**end while**

## VI. EXPERIMENTAL RESULTS

In this section, we first evaluate the network-adaptivity of our window-based algorithm, in the absence of any user interaction. Following that, we evaluate the user-adaptivity of our algorithm.

A. Network-adaptivity evaluation

We compare our window-based algorithm (WBA) with throughput-based and buffer-based algorithms (TBA and BBA) by stripping out of our algorithm the ability to deal with any user interaction, including rate-utility optimization. What is left is an algorithm that at each request opportunity $t_i$ puts just-received content at the end of the queue, and requests new content whose media time $\tau$ is expected to be newly uncovered by the leading edge of the window at time $t_{i+1}$, that is, $W_{lead}(t_i) < \tau \leq W_{lead}(t_i + T)$. The bandwidth $B_i$ of the representation of the content requested at time $t_i$ is the highest bandwidth available less than

$$\frac{C_i T}{W_{lead}(t_i + T) - W_{lead}(t_i)}. \quad (6)$$

where $C_i$ is the network throughput estimated at time $t_i$. Thus the client will request more data when the estimated throughput is high and the current buffer pointer $W_{lead}(t_i)$ is close to the target buffer pointer $W_{lead}(t_i + T)$. The buffer occupancy, which is the amount of media time in the queue, at any given time $t$ is $W_{lead}(t) - W_{trail}(t)$ in this case.

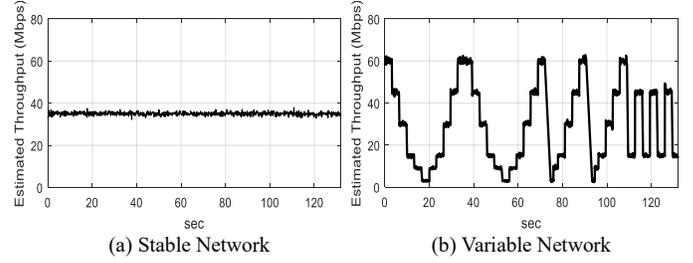

(a) Stable Network    (b) Variable Network

Fig. 9. Network model

All three algorithms (WBA, TBA, and BBA) are tested under the same network conditions, which are a combination of random and deterministic processes. Specifically, the packet arrival process is modeled as a Poisson random process whose mean value is deterministically controlled to model environmental factors such as service providers, traffic patterns, and wireless technologies. The throughput of wireless networks, for example, depends on the channel conditions and density of users. In this paper, we consider two different scenarios. The first scenario (Fig. 9(a)) is a stable network in which the mean packet arrival rate is constant. If the client is connected to the Internet through a wired network and there are only a few users, the network throughput may be very stable. The second scenario (Fig. 9(b)) is a variable network in which the mean packet arrival rate changes frequently. If the client is connected to the Internet through a wireless network and there are many users within a small area competing for the same wireless network resources, the network throughput may be very unstable.

Fig 10. shows simulation results for the stable network condition. The proposed algorithm (WBA) is compared with the throughput-based algorithm (TBA) and the buffer-based algorithm (BBA). The first row shows the buffer occupancy in GOFs and the second row shows the bandwidth selection results. There are 5 representations on the server, at bandwidths C1, C2, C3, C4, and C5 (4Mbps, 8Mbps, 12Mbps, 16Mbps, and 20Mbps respectively). The buffer occupancy of TBA increases continuously because it selects a representation with slightly lower bitrate than the estimated throughput. However, it neglects the buffer status, and it make the buffer occupancy increase, which means that it forfeits the chance to select better representations after the buffer becomes large. In contrast, the buffer occupancy of BBA stops growing and it chooses better

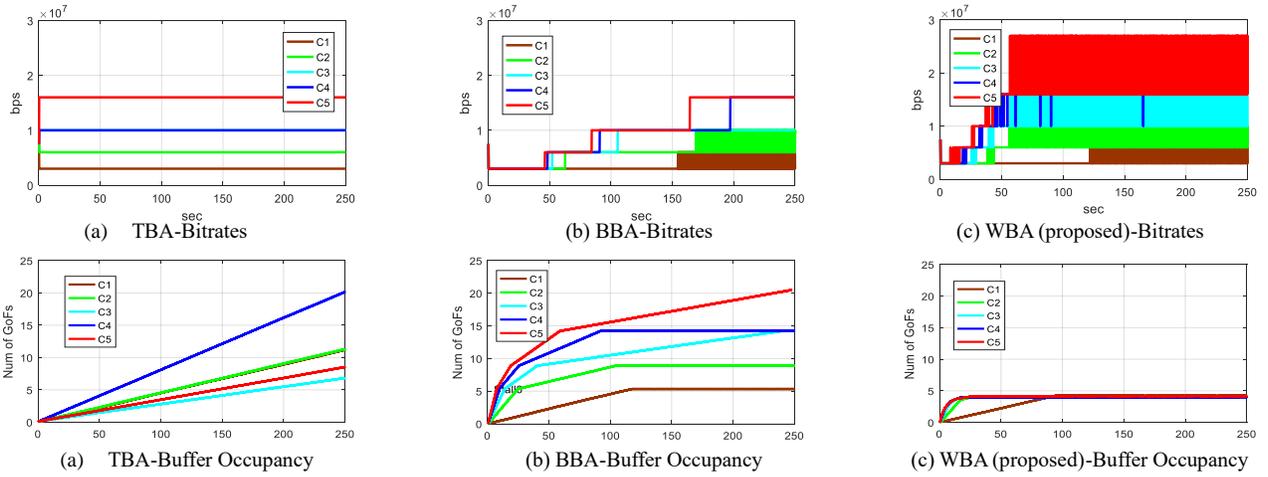

Fig 10. Stable Network Condition

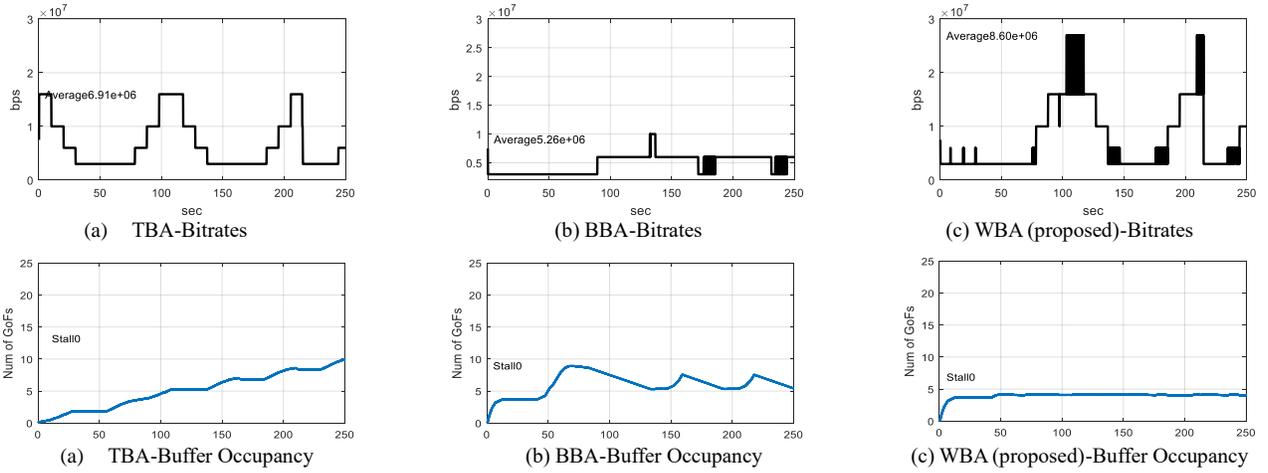

Fig 11. Variable Network Condition

representations. When the buffer occupancy is large, it drives the client to request better representations with higher bandwidths. The proposed algorithm shows a pattern similar to that of BBA, but it has better average throughput than both TBA and BBA, because it considers the estimated throughput and buffer occupancy at the same time.

Fig 11. shows simulation results for the variable network condition. The proposed algorithm again achieves the highest average bitrate (8.6 Mbps compared to 6.91 Mbps and 5.26 Mbps), and furthermore there are no stall (buffer underflow or rebuffering) events. TBA and BBA also do not have any buffer underflow events, because TBA chooses bitrates that are smaller than the estimated throughput, and BBA tries to increase the buffer occupancy rather than receiving higher bitrates. BBA needs a much larger buffer occupancy than the proposed method to prevent rebuffering events.

The point of this study is not to show that the proposed WBA is superior to the commonly used TBA or BBA, but rather to show that WBA is state-of-the-art from a network adaptation point of view, and that it does not give up any ability to be network-adaptive while also being user-adaptive.

B. User-adaptivity evaluation

We next investigate the performance of the proposed method when users are moving or controlling the viewpoint in AR applications. Deterministically generated camera paths are used to model user behavior. The camera path-1 (Fig. 12) starts far from the object and get close to the object, which models the zoom-in and zoom-out. The camera path-2 (Fig. 13) moves around an object seeing the object. This is for modeling the rotation of an object in a client's display. As camera move close to the object, smaller number of tiles is visible and those tiles should have higher quality.

To see the advantages of using the proposed algorithm for tile-based streaming, utility values are measured with different tile sizes. Fig. 14. shows simulation results for the stable network condition, and Fig 15. shows simulation results for the variable network condition. Depth indicates the number of dyadic divisions of cube to form tiles. With smaller tile sizes, we can achieve better utility in both cases, since we can allocate more bits to only visible tiles. Simulation results with a moving camera are provided to show the visibility adaptation performance of the proposed algorithm. We perform the simulations with two different speeds of camera movements to model slow and fast interactions.

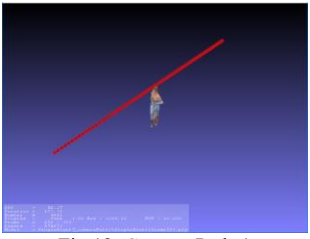 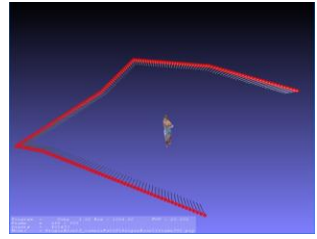

Fig 12. Camera Path-1    Fig 13. Camera Path-2

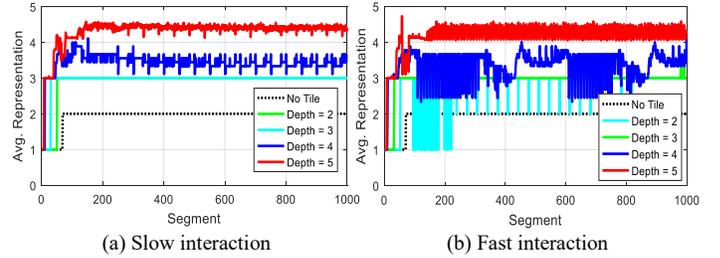

(a) Slow interaction    (b) Fast interaction

Fig. 14. Stable Network condition

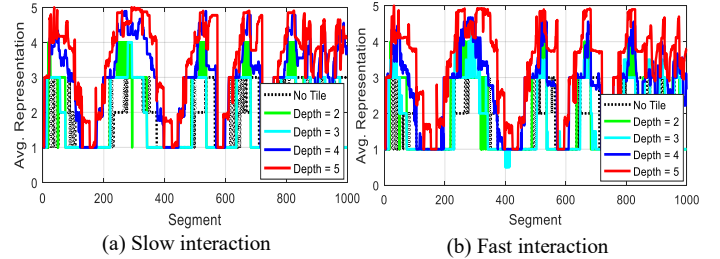

(a) Slow interaction    (b) Fast interaction

Fig. 15. Variable Network condition

Simulation results show that the algorithm chooses higher bitrates for visible tiles. In the variable network condition, sometimes network throughput gets very bad, but the window-based algorithm can tolerate the low throughput using existing data in the buffer and requesting lower bitrates to save bandwidth. The buffer management algorithm takes advantage of GOPs that already exist in the buffer when the video clip is repeated, re-using such GOPs without requesting new data, or alternatively upgrading the GOP to a higher quality representation. Using this feature, the client can continuously improve the quality of video. We can also see that the speed of camera movement does not affect the average representation level of the visible parts. The average representation level of the visible parts is hardly degraded when the camera moves faster when smaller tiles are applied. However, we still can achieve better average representation level of the visible parts by using smaller tiles, because it helps to reduce the redundant data streaming.

The proposed algorithm can be used for multiple objects without changing the algorithm. To illustrate, we run a simulation with five objects. Fig. 16 shows the location of objects, and the camera path. The multiple objects compete for limited bandwidth. Fig. 17. shows the estimated throughput, and the utility values of each object. The algorithm allocates the more bits to visible tiles of visible objects, since utility is high only when they are visible. However, the AR streaming system is always ready for a sudden change of user behavior. Therefore, the algorithm also requests non-visible tiles with lower quality to ensure that some data exists out of the user's view. Moreover, as it repeats the video, the user receives more tiles with better quality.

## VII. CONCLUSION

This is the first paper, to our knowledge, to deal with on-demand and/or live streaming of volumetric media for VR/AR applications. The main new aspect of streaming volumetric content for VR/AR applications, compared to streaming video (or even spherical video), is the high amount of user interactivity. This requires the streaming to be not only network-adaptive but also user-adaptive. Thus, new approaches are needed to minimize both network bandwidth and perceived latency due to user interaction.

We introduce window-based buffering to address the problem of minimizing latency after user interaction. Window-based buffering, in contrast to queue-based buffering, allows data to be requested for any part of the window, not just at the end of the queue. Fundamentally, this makes it possible to adapt quickly to unexpected user interactions while still being robust to variations in network throughput.

Window-based buffering has been used before, in the early days of streaming, for packet-level requests. This proved complicated, essentially because the round trip time (RTT) for a packet request dominated its download time, and packet delivery was not guaranteed. The uncertainty of having many unreliable packets in the air simultaneously required sophisticated tools such as Markov decision processes to optimize the packet requests. In contrast, in our system we request relatively large chunks of data over HTTP, whose download time dominates the RTT and whose delivery is guaranteed. Hence the requests can optimized serially. We introduce a simple greedy yet provably optimal algorithm for maximizing the expected utility of each request under a rate constraint.

We introduce 3D tiling to concentrate network resources on areas of interest to the user. 2D tiling has been used for a similar purpose in spherical video streaming. However, 3D tiles must be handled differently than 2D tiles because 1) there are polynomially many more 3D tiles than 2D tiles and 2) whether a tile is occupied by content or not is content-dependent and hence varying. We address both problems by creating a compact index into the set of occupied tiles in every group of frames. A further issue is that 3D tiles, unlike 2D tiles, may be occluded, depending on the viewpoint of the user.

Finally, we develop a comprehensive utility measure for each tile that takes into account the bitrate (as a proxy for the quality) of its representation, the spatial resolution of its representation, whether the tile is in the user's view and facing the user, the distance to the user, the spatial resolution of the display, and the time from when it is requested to when it will be displayed. We present a simple user prediction model, and pose a new research direction on user prediction modeling.

Our solution to volumetric streaming is consistent with

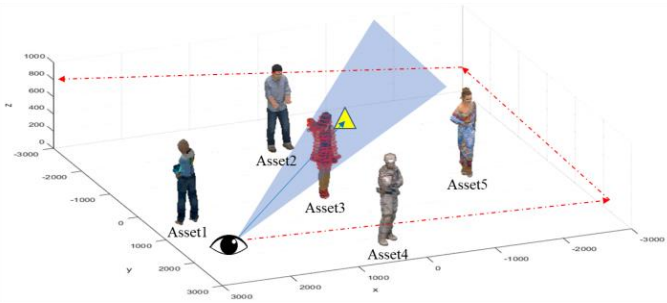

Fig 16. Multiple assets in the user's view

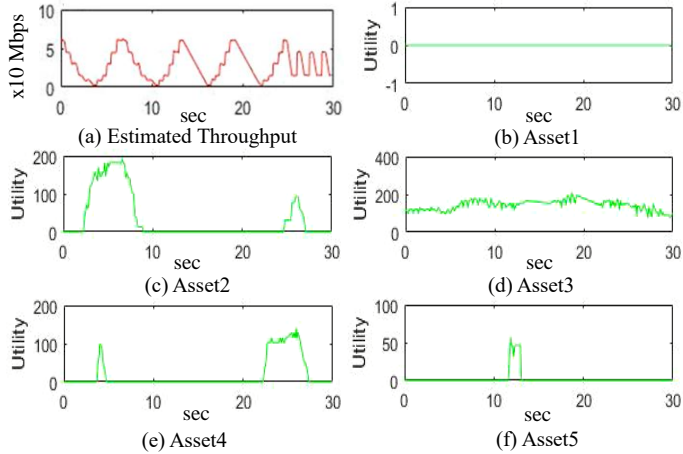

Fig 17. Utilities of Multiple Assets

modern HTTP streaming and requires only minor modifications to, for example, DASH, such as information in the manifest to support the notion of a volume and its coordinate system relative to the user, as well as tiles and segment indexes for the tiles. Throughout the paper, we provide details of how our solution supports features such as

- State-of-the-art network-adaptivity,
- Tile-based culling depending on predicted user frustum to save bandwidth,
- Tile-based level-of-detail depending on predicted user distance (and display resolution) to save bandwidth,
- Bandwidth allocation among multiple simultaneous objects,
- Sub-second response time to load new representations as necessary for user interaction,
- Fast start,
- Seek and fast-forward trick modes, and
- Quality improvement on subsequent replays if the data are stored.

Simulation results show that even in the absence of tiles and rate-utility optimization, our algorithm has both higher throughput and fewer rebuffering events than existing algorithms, in both stable and variable network conditions. Simulation results also show that our algorithm handles user interaction for volumetric video as expected, with sub-second latency.

APPENDIX: PSEUDOCODE

```
// Initialize
downloadBytes = 0
downloadTime = 0
Foreach object

  // Get manifest
  Get object.manifest (assume one period, one adaptation set)
  downloadBytes += number of downloaded bytes
  downloadTime  += download time
  Object.startTime = object.manifest.startTime
  Object.duration = object.manifest.duration
  Object.endTime = object.startTime + object.duration
  Object.startSegment = object.manifest.segmentTemplate.startNumber
  Object.segmentDuration = object.manifest.segmentTemplate.duration
  Object.segmentCount = ceil(object.duration/object.segmentDuration)
  Object.endSegment = object.startSegment + object.segmentCount - 1

  // Select clip to play and speed (or get this from playback engine)
  Object.clipStartTime = object.startTime // default to start of
   object
  Object.clipEndTime = object.endTime // default to end of object
  Object.clipSpeed = 1 // default to playback at 1x media time

  // Read first segment
  StartSegment = floor((object.clipStartTime-
   object.startTime)/object.segmentDuration)
  Get object.index[startSegment] from
   object.manifest.segmentTemplate.media
  downloadBytes += number of downloaded bytes
  downloadTime  += download time
  Object.minSegmentInWindow = startSegment
  Object.maxSegmentInWindow = startSegment
  Object.minTimeInWindow = object.startTime +
   (object.minSegmentInWindow - startSegment) *
   object.segmentDuration
  Object.maxTimeInWindow = object.startTime +
   (object.maxSegmentInWindow - startSegment + 1) *
   object.segmentDuration
  Object.currentTime = object.clipStartTime

// Estimate initial network rate
Chat = 8*downloadBytes / downloadTime
SmoothedChat = Chat
T = target time between requests // ideally should be interval
between playback of integer number of GOFs in steady state

// Loop until done
Playback.startTime = 0;
Playback.currentTime = 0;
Playback.endTime = max over objects ((object.clipEndTime -
 object.clipStartTime)/object.clipSpeed)
While playback.currentTime < playback.endTime

  // Determine desired window size (in time)
  Window.TargetDuration = max{5, playback.CurrentTime+1}

  // Get Index for all Segments covering Window
  downloadBytes = 0
  downloadTime = 0
  Foreach object
  While (object.maxTimeInWindow-
   object.clipStartTime)/object.clipSpeed <
   playback.currentTime+window.TargetDuration & object.startSegment +
   object.maxSegmentInWindow < object.endSegment
  Seg = ++object.maxSegmentInWindow
  Get object.index[seg] from object.manifest.segmentTemplate.media
  downloadBytes  += number of downloaded bytes
  downloadTime += download time
  Object.maxTimeInWindow = object.startTime +
   (object.maxSegmentInWindow + 1) * object.segmentDuration

  // Get user's current view in world coordinates from playback
   engine
  View[v].userToWorld
```

```
      View[v].frustum

// For each tile in window, compute utility and bitrate for each
 representation
Foreach object
  For seg = object.minSegmentInWindow to object.maxSegmentInWindow
    Foreach gof in object.index[seg]
      If (gof.startTime-object.clipStartTime)/object.clipSpeed <
       playback.currentTime || (gof.startTime-
       object.clipStartTime)/object.clipSpeed >
       playback.currentTime + window.targetDuration
        Continue
      P_err = 0.1+0.3*(min{1,((gof.startTime-
       object.clipStartTime)/object.clipSpeed -
       playback.currentTime)/5} // P_err of view predictor
      Foreach tile in gof
        Tile.(x,y,z) =
         object.cubeToWorld(mortonToXyz(tile.mortonCode))
        For each view v
          dist[v] = distance(tile.(x,y,z),
           view[v].userToWorld.(x,y,z)) // in meters
          P[v] = (tile.(x,y,z) in view[v].frustum &&
           dot(view[v].userToWorld.(x,y,z)-tile.(x,y,z),
           tile.normal) > 0)? (1-P_err) : P_err
        For n = 1 to object.representationCount
          Tile.bitCount[n] =
           object.representation[n].bandwidth/object.representation
           [n].framerate * gof.frameCount / seq.gof.tileCount
          Tile.utility[n] = max over v of
           alpha*log(beta*object.representation[n].bandwidth)*[(obj
           ect.tileWidth/object.maxWidth*fovOfCubeAt1m
           /dist[v])*min{object.representation[n].width/fovOfCubeAt
           1m*dist[v],display.horzPixels/view[v].frustum.horzFOV}]
           ^2 * P[v]
        Tile.maxLambda = max over m s.t. tile.bitCount[m]>
         tile.bitCount[tile.n] (tile.utility[m]-
         tile.utility[tile.n])/(tile.bitCount[m]-
         tile.bitCount[tile.n])
        Tile.argmaxLambda = argmax over m s.t. tile.bitCount[m]>
         tile.bitCount[tile.n] (tile.utility[m]-
         tile.utility[tile.n])/(tile.bitCount[m]-
         tile.bitCount[tile.n])

// Greedily allocate bits
R = 8*downloadBytes // bits already downloaded in index
While R < smoothedChat * T
  Tile = tile in buffer with maximum tile.maxLambda
  If tile.maxLambda <= 0 then break
  R += tile.bitCount[tile.argmaxLambda] - tile.bitCount[tile.n]
  Tile.n = tile.argmaxLambda
  If tile.n = argmax over m tile.bitCount[m]
    Tile.maxLambda = 0
    Tile.argmaxLambda = tile.n
  else
    Tile.maxLambda = max over m s.t. tile.bitCount[m]>
     tile.bitCount[tile.n] (tile.utility[m]-
     tile.utility[tile.n])/(tile.bitCount[m]-tile.bitCount[tile.n])
    Tile.argmaxLambda = argmax over m s.t. tile.bitCount[m]>
     tile.bitCount[tile.n] (tile.utility[m]-
     tile.utility[tile.n])/(tile.bitCount[m]-tile.bitCount[tile.n])

// Make request
Foreach object
  For seg = object.minSegmentInWindow to object.maxSegmentInWindow
    Initialize multi-part byte range requests
    Foreach gof in object.index[seg]
      If (gof.startTime-object.clipStartTime)/object.clipSpeed <
       playback.currentTime || (gof.startTime-
       object.clipStartTime)/object.clipSpeed >
       playback.currentTime + buffer.targetDuration
        Continue
      Foreach tile in gof
        If tile.n > 0
          Add byte range tile.byteOffset to
           tile.byteOffset+tile.byteCount (combining with previous
           range if possible) to request
        Tile.utility[0] = tile.utility[tile.n]
        Tile.n = 0
    Issue multi-part byte range request to
     object.manifest.segmentTemplate.media
    downloadBytes  += number of downloaded bytes
    downloadTime += download time

// Estimate network rate
Chat = 8*downloadBytes / downloadTime
SmoothedChat = f*Chat + (1-f)*smoothedChat

Advance playback.currentTime.  If playback.currentTime passes 1.5s,
 notify playback engine that it's safe to begin playing
```